\documentclass[12pt]{article}
\usepackage[dvips]{graphicx}
\newcommand{\tauv}{\mbox{\boldmath$\tau$}}
\usepackage{epsfig}
\usepackage{epsf}
\usepackage{amsfonts}
\usepackage{amssymb}

\def \be {\begin{equation}}
\def \ee   {\end{equation}}
\def \bea {\begin{eqnarray}}
\def \eea   {\end{eqnarray}}

\begin{document}

\title{\vskip -70pt
\begin{flushright}
{\normalsize DAMTP--2006--112} \\
\end{flushright}
\vskip 60pt {\bf On the Spin of the $B=7$ Skyrmion} \vskip 20pt}
{\author{ Olga~V.~Manko {\thanks{Email:
O.V.Manko@damtp.cam.ac.uk}}
\vspace{.2cm} \\
and \vspace{.2cm}
\\  Nicholas~S.~Manton{\thanks{Email: N.S.Manton@damtp.cam.ac.uk}}}}
\date{}
\maketitle

\vspace{-0.5cm}

\begin{center}
\textsl{\large{Department of Applied Mathematics and Theoretical
Physics
\vspace{.2cm} \\
Wilberforce Road, Cambridge CB3 0WA, UK}} \\\vspace{0.7cm}

\large{November 2006}
\medskip
\end{center}

\vspace{.1cm}

\begin{abstract} \noindent
{We investigate the collective coordinate quantization of the icosahedrally
symmetric $B=7$ Skyrmion, which is known to have a ground state with
spin $\frac{7}{2}$ and isospin $\frac{1}{2}$. We find a particular quantum
state maximally preserving the symmetries of the classical solution,
and also present a novel relationship between the quantum state
and the rational map approximation to the classical solution. We also
investigate the allowed spin states if the icosahedral symmetry is
partially broken. Skyrme field configurations with $D_5$ residual
symmetry can be quantized with spin $\frac{3}{2}$, giving a realistic
model for the ground states of the $^7{\rm Li}/^7{\rm Be}$ isospin doublet}.
\end{abstract}

\newpage

\section{Introduction}
Skyrmions, first introduced in \cite{Skyrme}, are topological
solitons in three spatial dimensions which are
candidates for the description of nuclei, the baryon (or
nucleon) number $B$ being identified with the topological soliton
number. Partly with the help of the rational map approximation
\cite{Manton}, minimal energy Skyrmion solutions have been found
numerically for baryon numbers up to $B=22$ and beyond, and their
symmetries have been determined \cite{Braaten,Battye,Sutcliffe}. In these
calculations a zero pion mass was usually assumed, and although
some recent developments \cite{new paper2,new paper1} show
that this approximation is not fully justified, for smaller $B$ it
remains a good model, and we will be using it in what follows. All
minimal energy Skyrmions with $B>2$ have only discrete symmetries,
i.e. are invariant under a discrete group of combined rotations
and isorotations, and their baryon densities are localized around the
edges of some polyhedra. After quantization of the collective
rotational and isorotational degrees of freedom (and possibly some
vibrational modes), these polyhedral density distributions are generally
smoothed into a more spherical form which one hopes give a good
match with the experimental shapes of nuclei. A quite
successful analysis was carried out in
\cite{Kop,BraatCar,Carson,Walhout,Irwin,ya}, showing that if
one performs the collective coordinate quantization of Skyrmions
with baryon numbers $B=2, 3, 4, \mbox{or } 6$, imposing the
Finkelstein--Rubinstein (FR) constraints associated with
the symmetries of the classical configurations \cite{FR}, which encode
the requirement that each $B=1$ Skyrmion is quantized as a fermion, then
the minimal energy solutions will have the right
spin/parity and isospin properties to model the deuteron,
$^3{\rm H}/^3{\rm He}$, $^4{\rm He}$ and $^6{\rm Li}$ respectively.

The $B=7$ solution \cite{Battye}, which is considered in this
paper, is particularly symmetric, having icosahedral $Y_h$
symmetry, with the baryon density being localized around the edges
of a dodecahedron. The collective coordinate quantization has been
considered in detail by Irwin \cite{Irwin} and Krusch
\cite{Krusch}, with the help of the rational map approximation.
However the lowest allowed spin state obtained with this approach
is $J=\frac{7}{2}$ (with the isospin being $I=\frac{1}{2}$), which
gives rise to a disagreement with real nuclei. Experimentally,
$J=\frac{7}{2}$ appears as the second excited state of the $^7{\rm
Li}/^7{\rm Be}$ doublet, with an excitation energy of 4.6 MeV
(relatively low for such a high spin), whereas the ground state
has spin $J=\frac{3}{2}$ and the nearby first excited state has
spin $J=\frac{1}{2}$. This suggests that the $B=7$ dodecahedral
Skyrmion is too symmetric to describe the physical ground state,
and the icosahedral group should be partially broken to allow for
states with smaller spin. A deformed Skyrmion will have a larger
classical potential energy than the undeformed one, but could be
energetically preferred because it would be quantized with a lower
spin and hence have lower kinetic energy. The kinetic energy
associated with the $J=\frac{7}{2}$ state has never been
calculated, but is of order 10--20 MeV. A $J=\frac{3}{2}$ state
has kinetic energy of order 2--5 MeV, so the potential energy of
the deformed configuration might be up to 15 MeV, still allowing
for a lower total energy.

This paper is organized as follows. In Section 2 we review the
basic concepts of the Skyrme model and the rational map ansatz,
which we use in the later sections. Then in Section 3 we present
the rational map for the $B=7$ Skyrmion, in various orientations.
In Section 4 we review the quantization of the $B=7$ Skyrmion,
and, following \cite{ya}, show using a slightly modified approach
to the quantized spin $\frac{7}{2}$ state of the Skyrmion, that
the dodecahedral density can be substantially preserved even in
the quantum case. We also show that the rational map itself
contains useful information about the quantum state. In Section 5
we examine different choices for breaking icosahedral symmetry,
find new ground states with various spins, and find the
corresponding collective coordinate wavefunctions. In particular,
we show that if one breaks the $D_3$ subgroup of $Y_h$, while
still preserving $D_5$ symmetry, the ground state will be the
observed $(J=\frac{3}{2}, I=\frac{1}{2})$ state. Alternatively, if
the $D_5$ subgroup is broken, both $(J=\frac{3}{2},
I=\frac{1}{2})$ and $(J=\frac{1}{2}, I=\frac{1}{2})$ are allowed.

\section{Skyrmions and the rational map ansatz}

In dimensionless units the Skyrme model with zero pion mass has
Lagrangian \be L =\int\left \{\frac{1}{2}
\textrm{Tr} (\partial_{\mu}U\partial^{\mu}U^{\dagger}) +
\frac{1}{16}\textrm{Tr}([\partial_{\mu}UU^{\dagger},
\partial_{\nu}UU^{\dagger}][\partial^{\mu}UU^{\dagger},
\partial^{\nu}UU^{\dagger}]) \right\} d^3x \,,
\label{lagrangian} \ee where $U(t,\textbf{x})$ is an
$SU(2)$--valued scalar field, which can be expressed nonlinearly
in terms of the isospin triplet of massless pion fields, and satisfies the
boundary condition $U(\mathbf{x})\rightarrow 1_2$ as
$\mathbf{x}\rightarrow\infty$. This boundary condition implies
that $U$ can be regarded as a map $U:S^3\rightarrow S^3$, where
the domain $S^3$ is identified with
$\mathbb{R}^3\bigcup\{\infty\}$, and the target $S^3$ is the manifold
of $SU(2)$. The topological degree of the map
$U$ has the explicit representation \be
B=-\frac{1}{24\pi^2}\int\varepsilon_{ijk}\textrm{Tr}
(\partial_iUU^{\dagger}\partial_jUU^{\dagger}\partial_kUU^{\dagger}) \,
d^3x \,. \ee The
conservation of the topological invariant $B$ makes it possible to
identify the solutions of the Skyrme field equation with
classical nuclei, with $B$ standing for the
baryon number. The lowest energy static solutions of the model, for
each $B$, are called Skyrmions.

Rational maps, i.e. holomorphic maps from $S^2\rightarrow S^2$,
prove to give good approximations to Skyrmion solutions,
especially those with low baryon number. One identifies the domain
$S^2$ of the rational map with a sphere in $\mathbb{R}^3$ centred
at the origin, of indeterminate radius, and the target $S^2$ with
the unit sphere in the Lie algebra of $SU(2)$. $\mathbb{R}^3$ can
be given coordinates $(r,z)$, where $r$ denotes the radius and the
complex variable $z$ denotes $\tan\frac{\theta}{2} \, e^{i\phi}$
(the stereographic coordinate, with $\theta$ and $\phi$ the usual
polar angles). The rational map $R(z)$ is a ratio of polynomials
in $z$. Its value $R$, at any point, corresponds (by stereographic
projection) to the Cartesian unit vector \be
{\bf{n}}_R=\frac{1}{1+|R|^2}(R+\bar{R},i(\bar{R}-R),1-|R|^2) \,.
\ee The rational map ansatz for the Skyrme field, depending on
$R(z)$ and a radial profile function $f(r)$, is \be
U(r,z)=\exp(if(r){\bf{n}}_{R(z)} \cdot {\bf{\tauv}}) \,, \label{4}
\ee where ${\tauv}=(\tau_1, \tau_2, \tau_3)$ is the triplet of
Pauli matrices, and $f(r)$ satisfies $f(0)=\bf\pi$, $f(\infty)=0$.

The baryon number for the ansatz (\ref{4}) is given by \be
B=-\int\frac{f'}{2\pi^2}\left(\frac{\sin{f}}{r}\right)^2
\left(\frac{1+|z|^2}{1+|R|^2}\left|\frac{dR}{dz}\right|\right)^2
\frac{2i \, dzd\bar{z}}{(1+|z|^2)^2}\,r^2 \, dr \,, \label{6} \ee
where $2i \, dzd\bar{z}/(1+|z|^2)^2$ is equivalent to the usual
2--sphere area element $\sin{\theta} \, d\theta d\phi$. The
angular part of the integrand, \be \left(\frac{1+|z|^2}{1+|R|^2}
\left|\frac{dR}{dz}\right|\right)^2 \,, \label{6'} \ee multiplied
by the area form $2i \, dzd\bar{z}/(1+|z|^2)^2$, is precisely the
pull--back of the area form $2i \, dR \, d\bar{R}/(1+|R|^2)^2$ on
the target sphere of the rational map, so its integral is $4\pi$
times the degree $N$ of the map. Therefore, with our choice for
the boundary conditions of $f$, (\ref{6}) simplifies to \be
B=\frac{-2N}{\pi}\int_0^{\infty}f'\sin^2f \, dr=N \,; \ee in other
words, a rational map of degree $B$ gives a Skyrmion of baryon
number $B$.

An $SU(2)$ M\"obius transformation of $z$ corresponds to a
rotation in physical space; an $SU(2)$ M\"obius transformation of
$R$ (i.e. on the target $S^2$) corresponds to an isospin rotation.
Both are symmetries of the Skyrme model, and preserve the baryon
number and energy. It is the rotational and isorotational
collective coordinates that we need to quantize. (There is also
translational symmetry in the Skyrme model, but its quantization
is standard, leading to momentum eigenstates.)

An attractive feature of the rational map ansatz is that it leads
to a simple classical energy expression which can be separately
minimised with respect to the parameters of the rational map
$R(z)$ and the profile function $f(r)$ to obtain close
approximations to the numerical, exact Skyrmion solutions, and almost
always with the correct symmetries. For some small values of $B$, including
$B=7$, there is a unique rational map of the desired degree with the
correct symmetry, which also minimises the angular part of the energy.

After quantizing the Skyrme field, much of the interesting
information, including the symmetries of the quantum baryon
density, is encoded in its angular part which only depends on
the rational map; therefore the profile function $f$ will not be
of much interest for our purposes.

\section{Icosahedrally symmetric $B=7$ Skyrmion}

The minimal energy $B=7$ Skyrmion has a dodecahedral shape, with
holes in the baryon density at the centres of the faces \cite{Battye}. The
symmetry group is the icosahedral group $Y_h$, whose rotational
subgroup is generated by a $2\pi/5$ rotation about a face, and
a $2\pi/3$ rotation about a vertex attached to that face. Their
product generates a $\pi$ rotation about the midpoint of an edge
attached to the face.

The Skyrmion can be well approximated by the rational map
ansatz, and there is an essentially unique $Y_h$--symmetric map of degree 7,
depending only on the choice of
orientation in space and isospace. One can orient the Skyrmion, so
that one or other of the above symmetry generators is manifest, and
the map then takes the following concise forms.

1) $C_5$ symmetry manifest \cite{Manton}:

\be R(z)=\frac{7z^5+1}{z^2(z^5-7)} \,; \label{1} \ee

2) $C_2$ symmetry manifest \cite{Manton}:

\be R(z)=\frac{bz^6-7z^4-bz^2-1}{z(z^6+bz^4+7z^2-b)} \,, \quad
b=7/\sqrt{5}\label{2} \,.\ee

3) For manifest $C_3$ symmetry the corresponding rational map has
not previously been found, and we obtain it here.
Recall that the Wronskian of a rational map \be
R(z)=\frac{P(z)}{Q(z)} \ee is defined as \be
W(z)=P'(z)Q(z)-P(z)Q'(z) \,. \label{Wronskdefn}\ee
We can rotate the Wronskian defined for
one of the previous orientations, and use this
to find the required map.

Consider the rational map (\ref{1}), whose Wronskian is a multiple of \be
W(z)=z^{11}+11z^6-z. \label{wronskian} \ee The Wronskian of a
general, degree 7 rational map is a 12th order polynomial;
therefore in our case the zeros of (\ref{wronskian}) are defined
by $\{z: W(z)=0\}\bigcup\{z=\infty\}$, and they are situated at
the face centres of the dodecahedron, the points of zero baryon
density. One of these face centres is at $z=0$. We now seek a
rotation which moves one of the vertices of the dodecahedron to
$z=0$. From \cite{Coxeter} we know that (in a certain orientation)
the Cartesian unit vectors pointing to a
vertex and nearest face center are \be \mathbf{n}_v
 = \frac{1}{\sqrt{3}} (0,\tau ^{-1},\tau) \qquad
\mbox{and} \qquad
\mathbf{n}_{f}=\frac{1}{\sqrt{1+\tau^2}}\,(0,\tau,1) \,, \ee
respectively, with $\tau=(1+\sqrt{5})/2$ the golden ratio.
Therefore, the rotation angle is given by \be
\cos{\lambda}=\mathbf{n}_v \cdot \mathbf{n}_{f} =
\frac{\tau^2}{\sqrt{3(1+\tau^2)}} \,. \ee The corresponding
M\"obius transformation for a rotation by the above angle,
preserving the real $z$--axis, is \be
z\rightarrow\tilde{z}=\frac{z\cos{\frac{\lambda}{2}-\sin{\frac{\lambda}{2}}}}
{z\sin{\frac{\lambda}{2}}+\cos{\frac{\lambda}{2}}} \,.
\label{mobius}\ee Acting with (\ref{mobius}) on the Wronskian
(\ref{wronskian}) we again get an 11th order polynomial in the
numerator. We must multiply this by the factor
$z+\cot{\frac{\lambda}{2}}$ corresponding to the zero that has
rotated from $z=\infty$. The result is the 12th order polynomial
with manifest $C_3$ symmetry: \be
W(z)=z^{12}+11\sqrt{5}z^9-33z^6-11\sqrt{5}z^3+1 \,. \ee The
corresponding rational map $P(z)/Q(z)$, where one of $P(z)$ and
$Q(z)$ is of degree 7 and the other of degree 7 or less, can be
worked out by solving (\ref{Wronskdefn}). This is a system of 12
equations in 15 variables, so there is some ambiguity, but it
disappears once we require manifest $C_3$ symmetry; this leads to
the unique rational map \be
R(z)=\frac{7z^6+7\sqrt{5}z^3+2}{z(2z^6-7\sqrt{5}z^3+7)} \,.
\label{3} \ee

\section{Quantization}

\subsection{FR constraints}

We quantize the $B=7$ Skyrmion as a rigid body free to rotate in
space and isospace, and assume here that it has its undistorted
dodecahedral shape. The symmetries of the Skyrmion impose restrictions on the
allowed quantum states via the FR constraints \cite{FR}. For each
symmetry element given by a rotation by $\alpha$ in space accompanied by
an isospin rotation by $\beta$ in the target space, we have to impose
the following condition on the wave function $\Psi$:
\be \exp{(i\alpha\mathbf{n}\cdot\mathbf{L})}
\exp{(i\beta\mathbf{N}\cdot\mathbf{K})}\Psi=\chi_{FR}\Psi \,, \ee
$\mathbf{n}$ and $\mathbf{N}$ being the directions of the
rotation axes in space and isospace, respectively, and $\mathbf{L}$ and
$\mathbf{K}$ the spin and isospin operators with respect to body fixed
axes. As usual for a rigid body, the spin and isospin operators with respect
to axes fixed in space, $\mathbf{J}$ and
$\mathbf{I}$, are distinct, but the Casimirs are the same:
$\mathbf{J}^2 = \mathbf{L}^2$ and $\mathbf{I}^2 = \mathbf{K}^2$.

For a Skyrmion whose symmetry is
captured by the rational map ansatz, the factor $\chi_{FR}$, which
is $\pm 1$, can be neatly evaluated using a formula due to Krusch
\cite{Krusch}: \be
\chi_{FR}=(-1)^{\cal N} \qquad \mbox{where}\qquad
{\cal N}=B(B\alpha-\beta)/2\pi \,. \label{Krusch} \ee
The rational map itself determines the correct sign choice for
the directions of the rotation axes and hence the
signs of the angles of rotation \cite{Krusch}. Let
$z_{\pm \mathbf{n}}$ denote the stereographic coordinates corresponding to
$\pm \mathbf{n}$, and similarly $R_{\pm \mathbf{N}}$ those corresponding to
$\pm \mathbf{N}$. The symmetry of the rational map $R(z)$ implies that
$R(z_{-\mathbf{n}})$ is one of $R_{\pm \mathbf{N}}$. Having chosen
$\mathbf{n}$, one should choose $\mathbf{N}$ so that
$R(z_{-\mathbf{n}}) = R_{\mathbf{N}}$. There is another
ambiguity for odd baryon numbers: namely one might add $2\pi$ to
the rotation or isorotation angles. This will not affect the
overall result, though, as the additional minus sign associated with
a $2\pi$ rotation or isorotation will be compensated by the
extra minus sign coming from $\chi_{FR}$ according to the formula
(\ref{Krusch}).

Looking carefully at the rational maps (\ref{1}), (\ref{2}) and
(\ref{3}), with icosahedral symmetry in various orientations,
one sees that a $2\pi/5$, $\pi$ or
$2\pi/3$ rotation about the $x_3$--axis, accompanied by
an isorotation by $4\pi/5$, $\pi$ or $2\pi/3$, respectively, about
the third isospin axis (both axes pointing up),
leaves the rational map unchanged.
Therefore, \bea
&&\chi_{FR}=-1 \,, \quad \mbox{for the $C_5$ generator} \,, \\
\nonumber
&&\chi_{FR}=-1 \,,  \quad \mbox{for the $C_2$ generator} \,, \\
\nonumber &&\chi_{FR}=+1 \,, \quad \mbox{for the $C_3$ generator} \,.
\eea Note that the product of the first two generators gives the
third, and the FR sign factors give a representation of this. More
generally, imposing the FR constraints for these generators
extends consistently to the whole $Y_h$ symmetry group. In
\cite{Irwin} it was shown that these constraints force the ground
state to have isospin $I=\frac{1}{2}$, and spin $J=\frac{7}{2}$.
The wavefunction, corresponding to the manifestly $C_5$--symmetric
orientation specified by (\ref{1}), is \bea \left|
\Psi\right\rangle
&=&\left\{\sqrt{\frac{7}{10}}\left|\frac{7}{2},-\frac{3}{2}\right\rangle
-\sqrt{\frac{3}{10}}\left|\frac{7}{2},\frac{7}{2}\right\rangle\right\}\otimes
\left|\frac{1}{2},-\frac{1}{2}\right\rangle \label{wf1} \\
\nonumber &&\quad +
\left\{\sqrt{\frac{7}{10}}\left|\frac{7}{2},\frac{3}{2}\right\rangle
+\sqrt{\frac{3}{10}}\left|\frac{7}{2},-\frac{7}{2}\right\rangle\right\}\otimes
\left|\frac{1}{2},\frac{1}{2}\right\rangle \,. \eea Here, the
terms in braces are the spin parts of the wavefunction, and the
second entry, after the total spin $\frac{7}{2}$, is the spin
projection onto the third body axis. (Note that because of the
$C_5$ symmetry, these values differ by 5.) These spin parts are
tensored with the isospin parts, where the second entry is the
(apparently unobservable) projection on to the third ``body''
isoaxis. Here, we do not specify the projection of the total spin
on to the third space axis, as this is arbitrary. The projection
of isospin on to the third ``space'' isoaxis can be either
$\frac{1}{2}$ or $-\frac{1}{2}$, giving a $^7{\rm Be}$ or $^7{\rm
Li}$ state, and is also not specified.

More explicitly, the wavefunction (\ref{wf1}) can be expressed in terms of
Wigner functions of the rotational and isorotational Euler angles.
$\Psi$ is then the amplitude to find the Skyrmion in the
orientation with those Euler angles relative to the standard
orientation of the rational map, (\ref{1}).

Since we are also interested in having the $C_3$ symmetry
manifest, we have calculated the wavefunction when the
standard orientation is that specified by (\ref{3}). This is \bea
|\Psi\rangle&=& \left\{{-\frac{\sqrt{2}}{3}}
 \left|\frac{7}{2},\frac{7}{2}\right\rangle
 +\sqrt{\frac{7}{18}}\left|\frac{7}{2},\frac{1}{2}\right\rangle-
 \sqrt\frac{7}{18}\left|\frac{7}{2},-\frac{5}{2}\right\rangle\right\}\otimes
 \left|\frac{1}{2},-\frac{1}{2}\right\rangle
\label{wf2} \\ \nonumber
&& \quad + \left\{\sqrt{\frac{7}{18}}\left|\frac{7}{2},\frac{5}{2}\right\rangle
 +\sqrt{\frac{7}{18}}\left|\frac{7}{2},-\frac{1}{2}\right\rangle+
 \frac{\sqrt{2}}{3}\left|\frac{7}{2},-\frac{7}{2}\right\rangle\right\}\otimes
 \left|\frac{1}{2},\frac{1}{2}\right\rangle \,. \eea

We have found a novel way to verify the structures of the
wavefunctions (\ref{wf1}) or (\ref{wf2}). The coefficients are
essentially the same as occur in the rational map itself. Observe
that for $B=7$ the rational map has a numerator which is, in
general, a degree 7 polynomial in $z$, with eight coefficients. The
denominator is similar. Under a M\"obius transformation of $z$,
corresponding to a rotation, the numerator and denominator each
change, with the coefficients transforming according to the
8-dimensional representation of $SU(2)$. This is the spin
$\frac{7}{2}$ representation. Similarly, under an isorotational
M\"obius transformation, the numerator and denominator are mixed
by the fundamental isospin $\frac{1}{2}$ representation.
Therefore, under rotations and isorotations, the quantum
wavefunction with spin $\frac{7}{2}$ and isospin $\frac{1}{2}$
transforms just like the rational map. Furthermore, we require the quantum
wavefunction to have, with respect to body fixed axes, the
symmetries of the rational map, i.e. the wavefunction is unchanged
under each symmetry operation, up to an FR sign factor. Again,
the rational map itself has
precisely these properties. So, to get the wavefunction, we just
take the coefficients of the rational map, and identify them with
the coefficients of the wavefunction. The one further step is to
correctly identify and normalise the basis elements.

For this last step, we note that a general degree 7 rational map
may be written in the form \be
R(z)=\frac{\sum_{s=-\frac{7}{2}}^{s=\frac{7}{2}}{P_s \, z^s}}
{\sum_{s=-\frac{7}{2}}^{s=\frac{7}{2}}{Q_s \, z^s}}\label{new rm} \,, \ee
where $s$ takes half odd-integer values. Taking $z =
\tan\frac{\theta}{2} \, e^{i\phi}$, we can identify each monomial
in $z$ in (\ref{new rm}) with a Wigner function
$D^{\frac{7}{2}}_{-s,\frac{7}{2}}$, in which
the projection of the spin onto the third space axis is maximal.
More precisely \be D^{\frac{7}{2}}_{-s,\frac{7}{2}}(\chi,\theta,-\phi)
=e^{i\frac{7}{2}\chi}\left(\frac{\sin\theta}{2}\right)^{7/2}
(-1)^{\frac{7}{2} +s}\left(\frac{7!}{(\frac{7}{2} -s)!(\frac{7}{2}+s)!}
\right)^{1/2} \, z^s\,. \ee So if we multiply the
numerator and denominator of the rational map (\ref{new rm}) by the common
factor $e^{i\frac{7}{2}\chi}\left(\frac{\sin\theta}{2}\right)^{7/2}$,
then it becomes a ratio of sums of Wigner functions. The numerator and
denominator become the (body) isospin down and isospin up parts of
the wavefunction. For example, the powers $z^7$ and $z^2$ in the
denominator of the rational map (\ref{1}) correlate with the spin
projection terms $-\frac{7}{2}$ and $\frac{3}{2}$ in the
wavefunction (\ref{wf1}). Also the ratio of coefficients in the
rational map, $1: \, -7$, converts to the ratio of coefficients in the
wavefunction $1:\sqrt{7/3}$ because of the normalisation factors
of the Wigner functions
$D^{\frac{7}{2}}_{-{\frac{7}{2}},{\frac{7}{2}}}$ and
$D^{\frac{7}{2}}_{{\frac{3}{2}},{\frac{7}{2}}}$.

This new way to obtain wavefunctions directly from the rational
map gives a remarkable significance to the rational map
approximation to Skyrmions. It means that, sometimes, the rational map
gives not only an approximation to the classical field
configuration, but also directly encodes information about the quantum
state. Unfortunately, this happens rather rarely. It
only works for odd baryon number, and where the isospin is
$\frac{1}{2}$ and the spin is $\frac{B}{2}$ (the dimensions of
these $SU(2)$ representations being 2 and $B+1$). The only other
frequently occurring example is the ground state of the $B=1$
Skyrmion, with spin and isospin $\frac{1}{2}$.

\subsection{Classical and Quantum Baryon Density}

In \cite{ya} we have noted that there is some choice for the
quantum baryon density of a Skyrmion. For example, the shape of
the quantum state will depend on the spin projection on to the
third spatial axis, $m$, which is usually considered arbitrary.
Here we follow the same logic as in \cite{ya} and construct a
quantum state $\Psi$ of the $B=7$ icosahedrally symmetric Skyrmion
which retains as much as possible of the spatial symmetry of the
classical solution. We do this by taking a suitable linear
combination of states with different $m$. The quantum baryon
density is then found by averaging the classical baryon density
over the collective coordinates weighted with $|\Psi|^2$. In what
follows, we will restrict our calculations to the states
(\ref{wf1}), based on the rational map (\ref{1}). The states
(\ref{wf1}) can be rewritten in terms of Wigner functions: \be |
\Psi_m\rangle=|\psi_{m}\rangle\otimes
\left|\frac{1}{2},-\frac{1}{2}\right\rangle+|\chi_{m}\rangle\otimes
\left|\frac{1}{2},\frac{1}{2}\right\rangle, \ee where
\bea|\psi_{m}\rangle&=&\sqrt{\frac{7}{10}}D_{-\frac{3}{2}
\,,m}^{\frac{7}{2}}(\alpha, \beta,
\gamma)-\sqrt{\frac{3}{10}}D_{\frac{7}{2}
\,,m}^{\frac{7}{2}}(\alpha, \beta, \gamma) \,,\nonumber\\
|\chi_{m}\rangle&=&\sqrt{\frac{7}{10}}D_{\frac{3}{2}
\,,m}^{\frac{7}{2}} (\alpha, \beta,
\gamma)+\sqrt{\frac{3}{10}}D_{-\frac{7}{2}
  \,,m}^{\frac{7}{2}}(\alpha, \beta, \gamma) \,. \eea The spatial
spin projection $m$ is now explicit, and $\alpha, \beta, \gamma$ are
the spatial Euler angles. From the orthogonality of different spin
and isospin states it follows that the required wavefunction, i.e.
the one which is icosahedrally symmetric with respect to both
body-fixed and space-fixed axes, is the combination \bea |
\Psi\rangle&=&\left\{\sqrt{\frac{7}{10}}\,|\psi_{-\frac{3}{2}}\rangle-
\sqrt{\frac{3}{10}}\,|\psi_{\frac{7}{2}}\rangle\right\}\otimes
\left|\frac{1}{2},-\frac{1}{2}\right\rangle\nonumber
\\\quad &&\quad +\left\{\sqrt{\frac{7}{10}}\,|\chi_{\frac{3}{2}}\rangle+
\sqrt{\frac{3}{10}}\,|\chi_{-\frac{7}{2}}\rangle\right\}\otimes
\left|\frac{1}{2},\frac{1}{2}\right\rangle \,. \label{icosstate}
\eea Note that the combination of $m$-values that is needed is
just the same combination as occurs in (\ref{wf1}).

The quantum baryon density is defined as \be
\rho_\Psi({\bf{x}})=\int
\mathcal{B}(D(A)^{-1}{\bf{x}})|\Psi(A)|^2\sin{\beta} \, d\alpha \,
d\beta \, d\gamma \,, \label{quantdens} \ee where $\Psi(A)$ is the
normalised wavefunction. Here $A$ stands for
the $SU(2)$ matrix parametrised by Euler angles $\alpha$, $\beta$,
$\gamma$, and $D(A)$ for the $SO(3)$ matrix associated to $A$ via
\be D(A)_{ab}=\frac{1}{2}{\rm Tr}(\tau_aA\tau_bA^\dag) \,, \ee and
$\mathcal{B}(\bf{x})$ is the classical baryon density of the $B=7$
Skyrmion. The important part is the angular dependence of
$\mathcal{B}(\bf{x})$, obtained by evaluating (\ref{6'}) for the
rational map (\ref{1}): \be
\mathcal{B}=\frac{196|z|^2(1+|z|^2)^2(z^{10}+11z^5-1)
(\bar{z}^{10}+11\bar{z}^5-1)}
{(|z|^{14}+49|z|^{10}+49|z|^4+1-7(|z|^4-1)(z^5+\bar{z}^5))^2} \,.
\ee Expressed in terms of polar angles, \bea &&\mathcal{B} = 196
\, \tan^2 \frac{\theta}{2}\left(1+\tan^2
\frac{\theta}{2}\right)^2 \\
&\times&\frac{\tan^{20}\frac{\theta}{2} +
22\tan^{15}\frac{\theta}{2}\,\cos{5\phi}
-2\tan^{10}\frac{\theta}{2}\,\cos{10\phi}+
121\tan^{10}\frac{\theta}{2}-
22\tan^5\frac{\theta}{2}\,\cos{5\phi}+1}
{(\tan^{14}\frac{\theta}{2}+49\tan^{10}\frac{\theta}{2}
-14\tan^9\frac{\theta}{2}\,\cos{5\phi}+14\tan^5\frac{\theta}{2}\,\cos{5\phi}
+49\tan^4\frac{\theta}{2}+1)^2} \,. \nonumber \eea It is
convenient to expand $\mathcal{B}$ in terms of spherical
harmonics $Y_{lm}(\theta,\phi)$: \be
\mathcal{B}=\sum_{l,m}{c_{lm}Y_{lm}(\theta,\phi)} \,, \label{7}
\ee where because of the icosahedral symmetry the lowest values of
$l$ are $0,6$ and 10, and the values of $m$ are multiples of 5.
The infinite series is dominated by the first four terms: \be
\mathcal{B}=c_{00}Y_{00}+c_{6-5}Y_{6-5}+c_{60}Y_{60}+
c_{65}Y_{65}+ \cdots \,, \label{classB} \ee and all
higher terms contribute less than a $10\%$ correction. Because the
map (\ref{1}) has degree 7, the integral of $\mathcal{B}$ over the
sphere is $28\pi$, so $c_{00}=14\sqrt{\pi}$, and we find
numerically $c_{65}=-c_{6-5}\simeq-6.38$, $c_{60}\simeq 7.97$ (see
Fig.1). The
ratio $c_{60}/c_{65}$ may also be found analytically from the fact
that \be Z=c_{6-5}Y_{6-5}+c_{60}Y_{60}+ c_{65}Y_{65} \label{z} \ee
is $Y_h$--symmetric and should have equal values at all the
Wronskian points (i.e. the dodecahedral face centres). In the
orientation we are considering the zeros of the Wronskian are
$z=0$, $z=\infty$, and the solutions of $z^{10}+11z^5-1=0$.
Solving this equation for $z^5$, and then explicitly calculating
the fifth root we find that the real, non-trivial Wronskian points
are $z=(-1\pm\sqrt{5})/2$. At $z=(-1+\sqrt{5})/2$, $\cos\theta =
1/\sqrt{5}$ and $\phi=0$, and we calculate that $Y_{65}\cong
0.428$, $Y_{60}\cong-0.334$. On the other hand, at $z=0$,
$Y_{65}=0$ and $Y_{60}\cong-1.017$. Hence, from (\ref{z}) we find
that $c_{60}/c_{65}\cong -1.25$, which agrees with the numerical
determination.

Using the transformation properties of spherical harmonics under
rotations, \be Y_{lm}(\tilde{\theta},
\tilde{\phi})=\sum_kD_{mk}^l(A)^*Y_{lk}(\theta,\phi) \,, \qquad
({\rm no  \,\, sum \,\, on \,\,}l) \,, \ee we can determine the
rotated baryon density in the integrand of (\ref{quantdens}), and
then using the integrals of the Wigner functions \bea
\int{D^j_{ab}(A)D^{j'}_{cd}}(A)^*\sin{\beta} \,  d\alpha \, d\beta
\, d\gamma&=&\frac{8\pi^2}{2j+1}\delta^{jj'}\delta_{ac}\delta_{bd}
\,, \\\nonumber
\int{D^j_{ab}(A)D^{j'}_{cd}(A)D^{j''}_{ef}(A)}\sin{\beta} \,
d\alpha \, d\beta \, d\gamma&=&8\pi^2\left(
\begin{array}{cccc}
j&j'&j''\\
a&c&e
\end{array}\right)\left(
\begin{array}{cccc}
j&j'&j''\\
b&d&f
\end{array} \right)\,,
\label{Wignerid} \eea we find that the angular dependence of
$\rho_\Psi$, the baryon density in the quantum state
(\ref{icosstate}), is
\be
\rho_\Psi=c_{00}Y_{00}+
0.23\left(c_{6-5}Y_{6-5}+c_{60}Y_{60}+ c_{65}Y_{65}\right) \,. \ee This is a
closed expression, since the higher order terms in the sum
(\ref{classB}) all average out to zero, and it resembles the
classical density, although more dominated by the first term, as can
be seen from Fig.2.

\section{Breaking the Icosahedral symmetry }

The state with spin $\frac{7}{2}$ considered so far is not the
observed ground state of $^7{\rm Be}$ or $^7{\rm Li}$, the nuclei
which should be described by the $B=7$ Skyrmion. We know from
experiment that the ground states form an isospin doublet with
spin $\frac{3}{2}$, and there is a nearby excited state with spin
$\frac{1}{2}$. It is encouraging that at energy only 4.6 MeV above the
ground state
there is a spin $\frac{7}{2}$ state, as this is not a high energy
for such a large spin. Nevertheless we still have the problem of
understanding the lower energy, lower spin states. One way to proceed
is to break some of the symmetries, thus
allowing more spin states. We seek a deformed $B=7$ Skyrmion, with
a higher classical potential energy than the dodecahedral
solution, but where the kinetic energy associated with the spin is
lower. In what follows, we shall just investigate the symmetries
and allowed spins, without seriously searching for the state of
lowest total energy. We will only be considering small
perturbations of the dodecahedral Skyrmion in order still to
be able to apply the rational map ansatz, and we assume that the FR
constraint associated with any unbroken symmetry is as before.

1) $D_3$ symmetry preserved. $D_3$ is generated by $C_2$ and $C_3$
symmetries, the axes of which are orthogonal. We assume the Skyrmion
is in the orientation of the rational map (\ref{3})
with $C_3$ symmetry about the
$x_3$--axis and $C_2$ symmetry about the $x_2$--axis. The $C_5$ symmetry
about any face centre of the dodecahedron is broken, if, for example,
the coefficient $7\sqrt{5}$ in the numerator and denominator of (\ref{3})
is increased or decreased.

The FR constraints take the form
\bea e^{\frac{2\pi i}{3}(L_3+K_3)}
\left|\Psi\right\rangle&=&\left|\Psi\right\rangle, \nonumber\\
e^{\pi
i(L_2+K_2)}\left|\Psi\right\rangle&=&\left|\Psi\right\rangle \,,
\eea where $L_i$ and $K_i$ are generators of spin and isospin
rotations respectively. The lowest allowed state is
$J=\frac{1}{2}$, $I=\frac{1}{2}$, with wave function \be
\left|\Psi\right\rangle=\left|\frac{1}{2},
\frac{1}{2}\right\rangle\otimes\left|\frac{1}{2},-\frac{1}{2}\right\rangle
-\left|\frac{1}{2},
-\frac{1}{2}\right\rangle\otimes\left|\frac{1}{2},\frac{1}{2}\right\rangle
\,. \ee Spin $\frac{3}{2}$ is also allowed leading to \be
\left|\Psi\right\rangle=\left|\frac{3}{2},
\frac{1}{2}\right\rangle\otimes\left|\frac{1}{2},-\frac{1}{2}\right\rangle
+\left|\frac{3}{2},
-\frac{1}{2}\right\rangle\otimes\left|\frac{1}{2},\frac{1}{2}\right\rangle
\,. \ee

2) $D_5$ symmetry preserved. This is generated by $C_2$ and $C_5$
symmetries about orthogonal axes. Here, we assume the Skyrmion
is in the orientation of the rational map (\ref{1})
with $C_5$ symmetry about the
$x_3$--axis and $C_2$ symmetry about the $x_2$--axis. The $C_3$ symmetry
is broken if the coefficient $7$ in the numerator and denominator of (\ref{1})
is varied. The FR constraints are \bea
e^{\frac{2\pi i}{5}(L_3+2K_3)}\left|\Psi\right\rangle
=-\left|\Psi\right\rangle \,, \nonumber\\
e^{\pi i(L_2+K_2)}\left|\Psi\right\rangle=\left|\Psi\right\rangle
\,. \eea This is the most interesting case, as the lowest allowed
state with isospin $\frac{1}{2}$ has spin $\frac{3}{2}$. The
corresponding wave function is \be \left|
\Psi\right\rangle=\left|\frac{3}{2},
\frac{3}{2}\right\rangle\otimes\left|\frac{1}{2},\frac{1}{2}\right\rangle
+\left|\frac{3}{2},
-\frac{3}{2}\right\rangle\otimes\left|\frac{1}{2},-\frac{1}{2}\right\rangle
\,. \ee

3) The degree 7 rational map \be
R(z)=\frac{bz^6-7z^4-bz^2-1}{z(z^6+bz^4+7z^2-b)} \,, \ee has
icosahedral symmetry when $b = \pm 7/\sqrt{5}$, but for general real
values of $b$ there is only tetrahedral $T_h$ symmetry. This symmetry
is generated by a $\pi$
rotation about the $x_3$--axis and a $2\pi/3$ rotation permuting
the Cartesian axes. Therefore, we obtain the following FR
constraints: \bea e^{\frac{2\pi
i}{3\sqrt{3}}(L_1+L_2+L_3)}e^{-\frac{2\pi
i}{3\sqrt{3}}(K_1+K_2+K_3)}|\Psi\rangle&=&|\Psi\rangle \,, \nonumber\\
e^{\pi i(L_3+K_3)}|\Psi\rangle&=&-|\Psi\rangle \,. \eea This time
the lowest allowed state is $J=\frac{1}{2}$, $I=\frac{1}{2}$, and
the corresponding wavefunction is \be \left|
\Psi\right\rangle=\left|\frac{1}{2},
\frac{1}{2}\right\rangle\otimes\left|\frac{1}{2},\frac{1}{2}\right\rangle
+i\left|\frac{1}{2},
-\frac{1}{2}\right\rangle\otimes\left|\frac{1}{2},-\frac{1}{2}\right\rangle
\,. \label{tetrahedral rational map} \ee There is no
$J=\frac{3}{2}$, $I=\frac{1}{2}$ state.

A further special case is when $b=0$, since the rational map
\be R(z)=-\frac{7z^4+1}{z^3(z^4+7)}
\label{cubic rational map} \ee has cubic
symmetry. The cubic group is
generated by a $2\pi/3$ rotation cyclically permuting the
Cartesian axes and a $\pi/2$ rotation about the $x_3$--axis.
Applying these to (\ref{cubic rational map}), one finds the
accompanying isospin rotations. The corresponding FR constraints
are \bea e^{\frac{2\pi i}{3\sqrt{3}}(L_1+L_2+L_3)}e^{-\frac{2\pi
i}{3\sqrt{3}}(K_1+K_2+K_3)}|\Psi\rangle&=&|\Psi\rangle \,, \nonumber\\
e^{\frac{\pi i}{2}L_3}e^{\frac{3\pi
i}{2}K_3}|\Psi\rangle&=&-|\Psi\rangle \,. \eea One can check that
the wavefunction (\ref{tetrahedral rational map}) satisfies these
constraints. Therefore spin $\frac{1}{2}$ is again allowed.

To find the baryon density corresponding to the most interesting case
2) above, let us perturb the rational map (\ref{1}) to
\be
R(z)=\frac{7z^5+1+a}{z^2((1+a)z^5-7)} \,, \ee which is still
$D_5$--symmetric. Then, to linear order in $a$,
the classical baryon density gets an additional contribution \bea
&\mathcal{B}&=196a|z|^2(1+|z|^2)^2
\nonumber\\
&& \, \times \left\{\frac{(z^{10}+11z^5-1)
(\bar{z}^{10}+\bar{z}^5-1)+(z^{10}+z^5-1)
(\bar{z}^{10}+11\bar{z}^5-1)}
{(|z|^{14}+49|z|^{10}+49|z|^4+1-7(|z|^4-1)(z^5+\bar{z}^5))^2} \right.\\
&& \, -\left.2\,\frac{(z^{10}+11z^5-1)
(\bar{z}^{10}+11\bar{z}^5-1)
(2|z|^{14}+2-7(|z|^4-1)(z^5+\bar{z}^5))}
{(|z|^{14}+49|z|^{10}+49|z|^4+1-7(|z|^4-1)(z^5+\bar{z}^5))^3}\right\}
\,.\nonumber \eea This gives rise to new, non-icosahedrally
symmetric terms in the spherical harmonic expansion \be
\mathcal{B}=\sum_{l,m}{c_{lm}Y_{lm}(\theta,\phi)} \,. \ee The
terms which give $90\%$ of the new contribution are $c_{20}\simeq
-12a$, $c_{40}\simeq-4a$ and the corrections to $c_{60}$ and
$c_{65}$ which are $\Delta c_{60}\simeq 5a$ and $\Delta
c_{65}\simeq-4.7a$. Since we are considering small $a$, the change
in the classical baryon density is small. When performing the
quantization, the situation changes dramatically. Now we are
working with a $J=\frac{3}{2}$ wave function. The series for the
quantum baryon density will be finite: \be
\rho_{\Psi}=c_{00}Y_{00}+0.25c_{20}Y_{20} \,. \ee Thus, we have a
very slightly deformed spherically symmetric density distribution.
The density, for $a = 0.1$, is shown in Fig.3.

\section{Conclusion}
In this paper we have considered a fundamentally new approach to
the quantization of the $B=7$ Skyrmion. We have shown that the
problem of inconsistency between the experimental ground state,
and the ground state coming from collective coordinate quantization of the
icosahedrally symmetric Skyrmion, might be overcome by breaking
part of the symmetry. We have focussed on three
different unbroken subgroups of $Y_h$, have found the corresponding lowest
allowed spin states, and the wavefunctions describing these.
For the case where the symmetry is broken to $D_5$,
the lowest allowed spin is $J=\frac{3}{2}$, so this type of
deformed Skyrmion is the best candidate for modelling the ground states
of the $^7{\rm Li}/^7{\rm Be}$ isospin doublet. Encouragingly,
Baskerville \cite{Bask} found that the lowest frequency,
parity-preserving, vibrational
mode of the $B=7$ Skyrmion is a squashing and stretching mode preserving $D_5$
symmetry. This means that for a given amplitude of deformation, the
extra potential energy is rather small. Other deformed Skyrmions,
retaining other symmetries, have both $J=\frac{1}{2}$
and $J=\frac{3}{2}$ spin states. It
would be interesting to actually determine the state with the
lowest total energy, but that requires a more quantitative investigation
of deformation energies and moments of inertia,
the topic for some future work.
In any case, we now have a promising approach to match the
properties of the real Lithium-7 and Beryllium-7 nuclei with
the results coming from the previously problematic $B=7$ sector of the
Skyrme model.

\newpage

\begin{figure}
\begin{center}
\mbox{\begin{picture}(250,350)(100,0)
\includegraphics[width=1\textwidth]{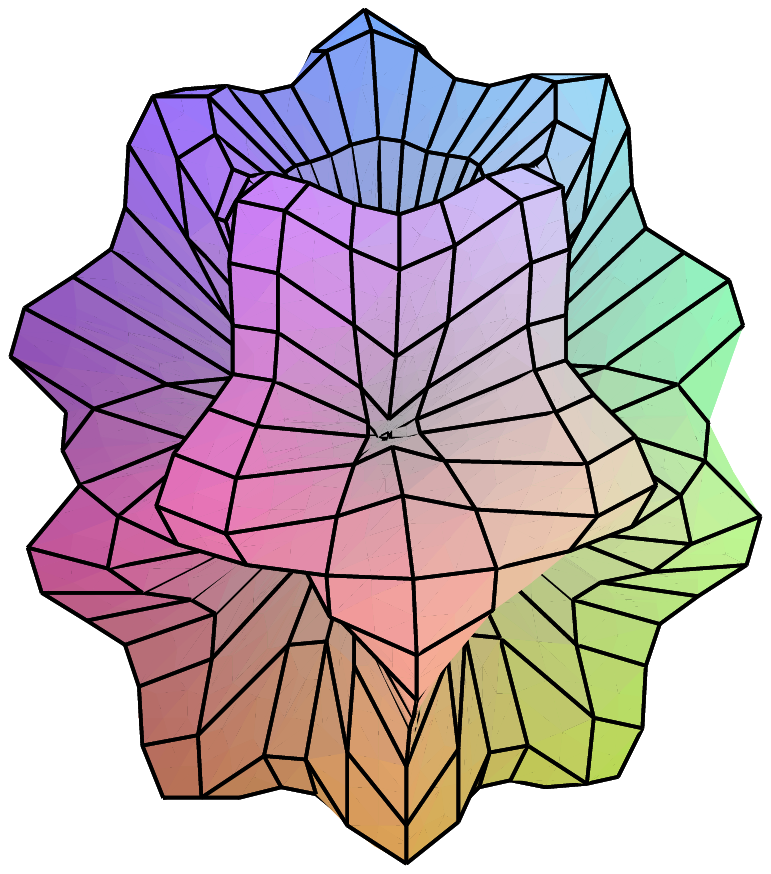}
\end{picture}}
\caption{Classical baryon density of $B=7$ Skyrmion,
truncated to first four terms in the expansion (33).}
\end{center}
\end{figure}

\newpage

\begin{figure}
\begin{center}
\mbox{\begin{picture}(250,350)(100,0)
\includegraphics[width=1\textwidth]{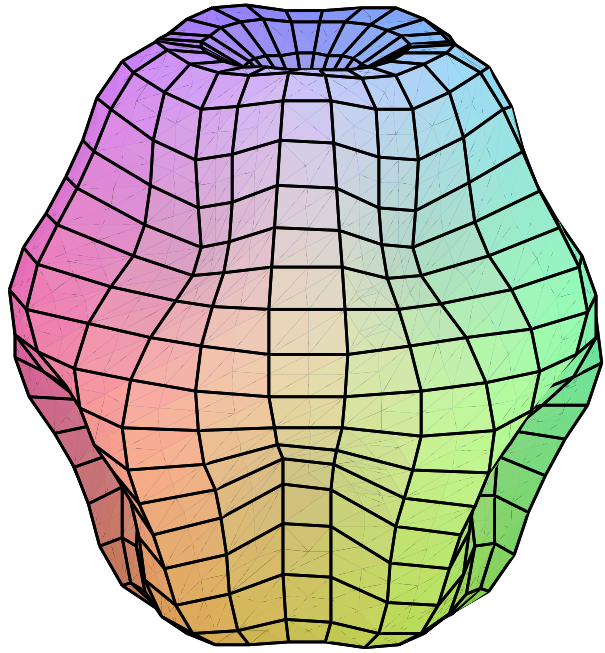}
\end{picture}}
\caption{Quantum baryon density of $B=7$ Skyrmion in spin $\frac{7}{2}$
state.}
\end{center}
\end{figure}

\newpage

\begin{figure}
\begin{center}
\mbox{\begin{picture}(250,350)(100,0)
\includegraphics[width=1\textwidth]{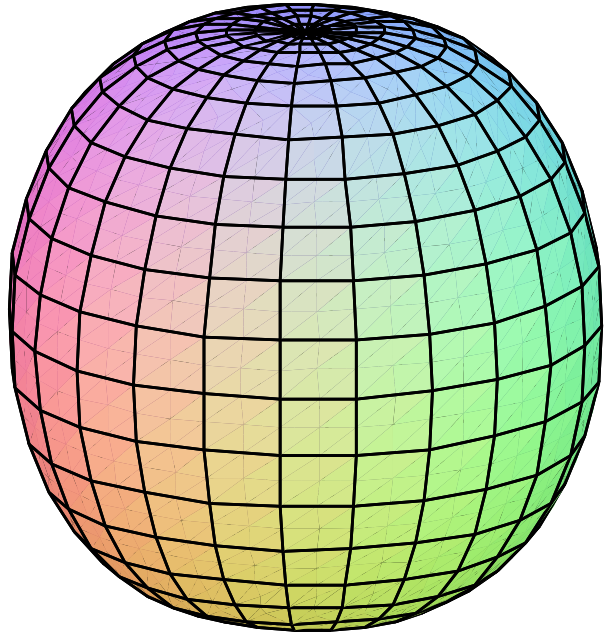}
\end{picture}}
\caption{Quantum baryon density of deformed Skyrmion with
residual $D_5$ symmetry, in spin $\frac{3}{2}$ state. Effect of deformation
$a$ is exaggerated 10 times.}
\end{center}
\end{figure}

\end{document}